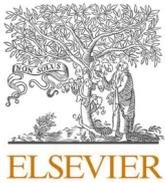
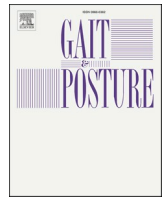

Short communication

# Between-session reliability of skin marker-derived spinal kinematics during functional activities

Luzia Anna Niggli [a,b], Patric Eichelberger [a], Christian Bangerter [a], Heiner Baur [a], Stefan Schmid [a,]*

[a] *Bern University of Applied Science, Department of Health Professions, Division of Physiotherapy, Spinal Movement Biomechanics Group, Bern, Switzerland*
[b] *ETH Zurich, Department of Health Science and Technology, Zurich, Switzerland*

ARTICLE INFO

ABSTRACT

*Background:* Skin marker-based analysis of functional spinal movement is a promising method for quantifying longitudinal effects of treatment interventions in patients with spinal pathologies. However, observed day-to-day changes might not necessarily be due to a treatment intervention, but can result from errors related to soft tissue artifacts, marker placement inaccuracies or biological day-to-day variability.
*Research question:* How reliable are skin marker-derived three-dimensional spinal kinematics during functional activities between two separate measurement sessions?
*Methods:* Twenty healthy adults (11 females/9 males) were invited to a movement analysis laboratory for two visits separated by 7–10 days. At each visit, they performed various functional activities (i.e. sitting, standing, walking, running, chair rising, box lifting and vertical jumping), while marker trajectories were recorded using a skin marker-based 10-camera optical motion capture system and used to calculate sagittal and frontal plane spinal curvature angles as well as transverse plane segmental rotational angles in the lumbar and thoracic regions. Between-session reliability for continuous data and discrete parameters was determined by analyzing systematic errors using one sample T-tests as well as by calculating intraclass correlation coefficients (ICCs) and minimal detectable changes (MDCs).
*Results and Significance:* The analysis indicated high relative consistency for sagittal plane curvature angles during all activities, but not for frontal and transverse plane angles during walking and running. MDCs were mostly below 15°, with relative values ranging between 10 % and 750 %. This study provides important information that can serve as a basis for researchers and clinicians aiming at investigating longitudinal effects of treatment interventions on spinal motion behavior in patients with spinal pathologies.

## 1. Introduction

Optical motion capturing is an appropriate method to quantify the biomechanics of the spine during functional movement [1]. It provides an important basis for the longitudinal evaluation of treatment effects in patients with spinal disorders. However, observed day-to-day changes might not necessarily be due to an administered treatment intervention, but can result from errors related to instrumental inaccuracies, soft tissue artifacts, marker placement inaccuracies or biological day-to-day variability [2–4]. While inaccuracies of current optical motion capture systems are relatively small (below 2 mm for dynamic experiments [5,6]), errors emerging from soft tissue artifacts and marker placement inaccuracies are considerably higher (up to 10.7 mm and 21.0 mm, respectively [7,8]). Most of these factors can only be partially controlled and therefore, it is important to know the extent of variability resulting from these factors. Researchers and clinicians need such information to determine whether an observed change can in fact be ascribed to a treatment intervention or is just a result of the aforementioned error sources. So far, a few studies reported day-to-day changes of trunk kinematics during functional activities [9–12], but since they only included evaluations of parameterized inter-segmental or inter-marker angles during walking, chair rising and box lifting, changes related to other functional activities such as running and jumping as well as angles at each time instance (i.e. continuous data) and clinically more meaningful measures such as spinal curvature angles remain unknown.

This study aims at evaluating the between-session reliability of three-dimensional spinal kinematics (continuous data as well as discrete parameters) during various functional activities derived from marker-based optical motion capturing.

## 2. Methods

### 2.1. Participants

Twenty healthy adults (11 females/9 males; height: 173 ± 10 (157–192) cm; mass: 69 ± 13 (45.5–91.7) kg; age: 31 ± 9 (20–53) years; body mass index (BMI): 22.6 ± 2.6 (18.5–27.5) kg/m$^2$) participated in this study. Recruitment took place via flyer and inquiries among the community surrounding the investigators' institution. The protocol was approved by the local ethics committee and written informed consent was obtained prior to the first measurement.

### 2.2. Measurement procedures and data collection

Participants were invited to the movement analysis laboratory for two visits separated by 7–10 days. At both visits, the same experienced physiotherapists equipped them with 58 retro-reflective markers according to a previously described configuration [1]. Participants were then asked to sit and stand quietly for 10 s and to perform four repetitions of the following activities (barefoot and at self-selected normal speed): walking and running on a 10-meter level ground, standing up and sitting down on a chair (chair rising), lifting up and putting down a 5 kg-box (box lifting), and performing a vertical counter movement jump (CMJ). Details on standardization and execution of the activities can be found elsewhere [13]. Data were recorded at a sampling frequency of 200 Hz using a 10-camera optical motion capture system (8x Bonita 3 and 2x Bonita 10; Vicon, Oxford, UK).

### 2.3. Data reduction

Following data pre-processing with the software Nexus (version 2.9.2., Vicon UK, Oxford, UK), we used custom MATLAB algorithms (R2019a, MathWorks Inc., Natrick, MA, USA) to calculate lumbar and thoracic sagittal plane curvature angles for all activities as well as frontal plane curvature angles and transverse plane segmental rotation angles for the activities walking and running. Detailed information on event detection, angle calculations, marker placement accuracy and soft tissue artefacts have been reported elsewhere [8,13–15]. In brief, curvature angles were calculated based on a circle fit algorithm applied to the markers placed on the spinous processes, whereas segmental rotation angles were determined by calculating the relative angles between the intersecting lines of marker pairs placed in the upper and lower parts of the lumbar and thoracic regions, respectively. Data were then low-pass filtered at 6 Hz (Butterworth, fourth order, zero-phase), time-normalized to cycles consisting of 101 frames and parameterized into average (only standing, sitting, walking and running activities) and range of motion (RoM) values (all dynamic activities).

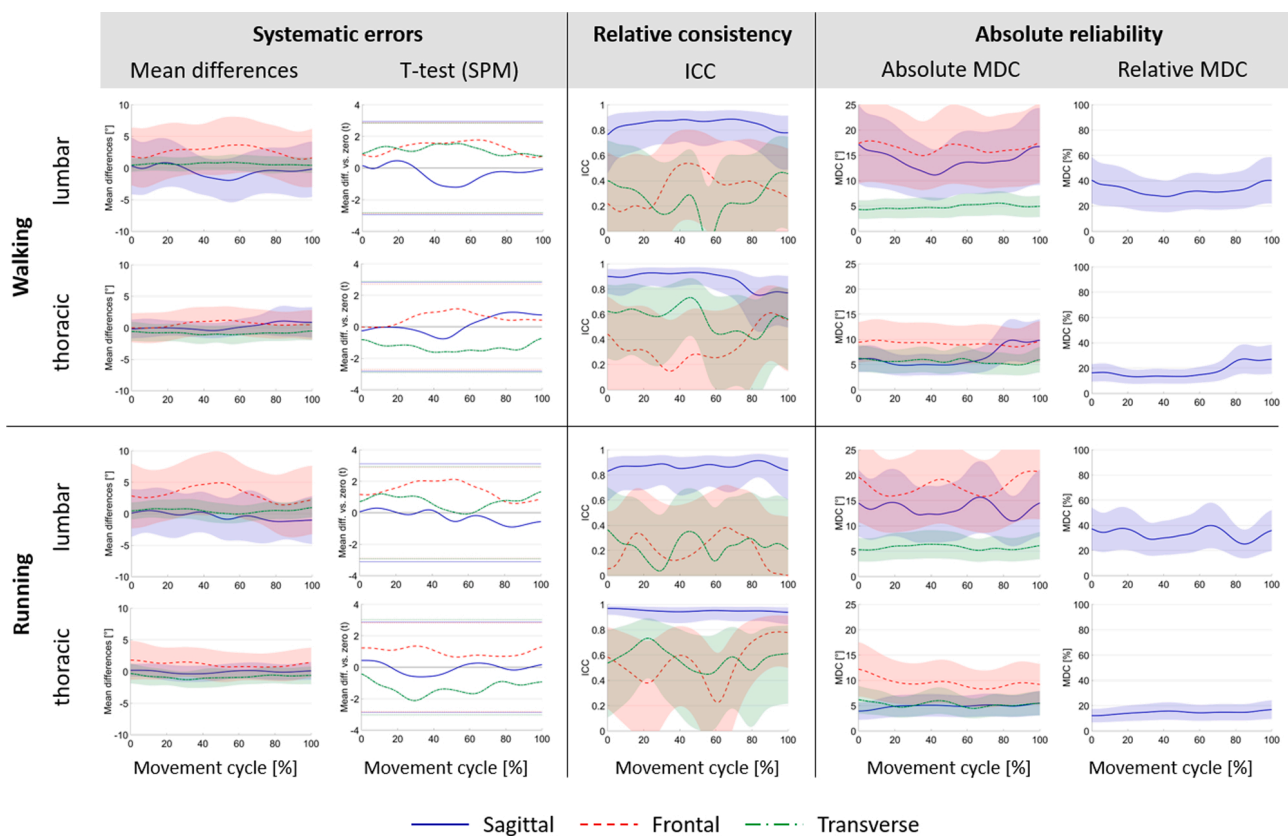

**Fig. 1.** Between-session reliability for continuous lumbar and thoracic spine angles in the sagittal (blue), frontal (red), and transverse (green) planes for the activities walking and running. The left column shows results of the evaluation for systematic errors using independent samples T-test (implemented by means of one-dimensional Statistical Parametric Mapping, SPM), with the horizontal colored lines indicating the thresholds for statistical significance at the p ≤ 0.05 level. Middle and right columns illustrate the intraclass correlation coefficient (consistency formula ICC(C,1)) for relative consistency and the minimal detectable change (MDC) for absolute reliability, respectively. The MDCs are thereby expressed as absolute values in degrees as well as relative values in percent of the mean value of the first and second measurement sessions. Due to multiple zero-crossings of the frontal and transverse plane angles, relative MDCs are therefore only presented for the sagittal plane. The shaded areas indicate 95 % confidence intervals (For interpretation of the references to colour in this figure legend, the reader is referred to the web version of this article).

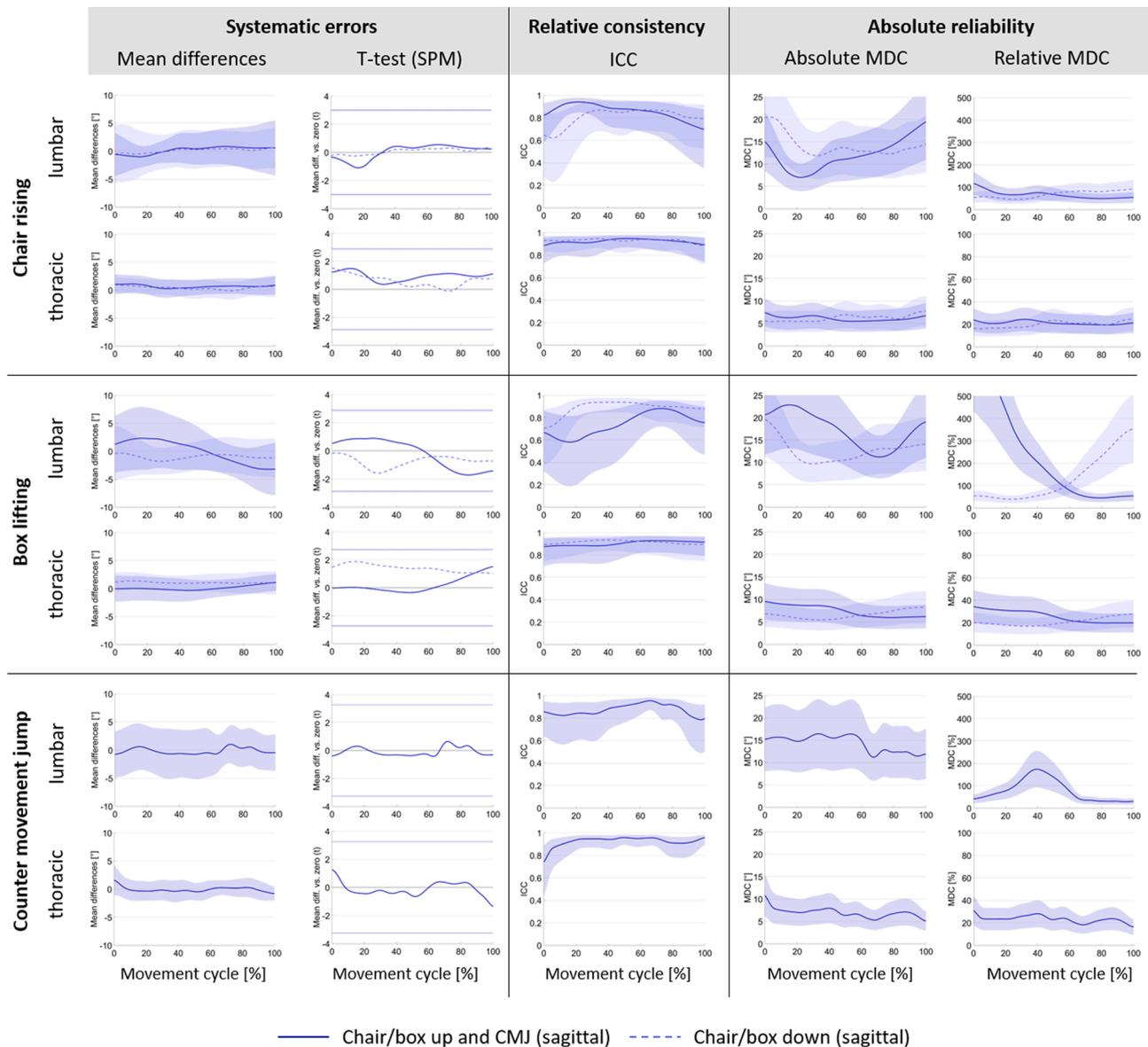

**Fig. 2.** Between-session reliability for continuous lumbar and thoracic sagittal plane spine angles for the activities chair rising (i.e. standing up (blue) and sitting down on a chair (red)), box lifting (i.e. lifting up (red) and putting down a box (blue)) as well as counter movement jump. The left column shows results of the evaluation for systematic errors using independent samples T-test (implemented by means of one-dimensional Statistical Parametric Mapping, SPM), with the horizontal colored lines indicating the thresholds for statistical significance at the $p \leq 0.05$ level. Middle and right columns illustrate the intraclass correlation coefficient (consistency formula ICC(C,1)) for relative consistency and the minimal detectable change (MDC) for absolute reliability, respectively. The MDCs are thereby expressed as absolute values in degrees as well as relative values in percent of the mean value of the first and second measurement sessions. The shaded areas indicate 95 % confidence intervals (For interpretation of the references to colour in this figure legend, the reader is referred to the web version of this article).

*2.4. Statistical analyses*

Statistical analyses were carried out using custom MATLAB algorithms (R2019a, MathWorks Inc., Natrick, MA, USA). Data collected from the functional activities were averaged over the four repetitions. Between-session reliability was evaluated based on the three-layered approach by Weir [16]. In a first step, data were thereby analyzed for systematic errors by using one sample T-tests (alpha-level: 0.05) to compare the mean differences to zero. For continuous data, T-tests were implemented using one-dimensional Statistical Parametric Mapping (SPM; spm1d-package www.spm1d.org), which applies Random Field Theory (RFT) to identify so called "supra-threshold clusters", i.e. clusters where the t-value for a comparison crosses a critical threshold that corresponds to a pre-defined alpha-level [13,17]. Subsequently, relative consistency was determined using intraclass correlations coefficients (consistency formula ICC(C,1)), and finally, absolute reliability was quantified by the minimal detectable change (MDC; a common index for determining the minimal difference required to be considered real), calculated as 1.96*SDd (standard deviation of the differences) [18] and expressed as a percentage of the mean values of both sessions.

**3. Results**

Some trials had to be excluded due to incomplete or missing marker data. A complete set of continuous and discrete angle data as well as additional information regarding reliability can be found in the Electronic Supplementary Material.

Table 1

Between-session reliability for average (only standing, sitting, walking and running activities) and range of motion (RoM) values (all dynamic activities) of lumbar and thoracic spine angles in the sagittal, frontal, and transverse planes for the activities standing, sitting, walking, running, standing up from a chair (chair up), sitting down on a chair (chair down), lifting up a box (box up), putting down a box (box down) and counter movement jump (CMJ). Presented are the results for systematic error tests (mean differences and t-tests), relative consistency (intraclass correlation coefficients (ICC)) and absolute reliability (minimal detectable change (MDC)). The MDCs are thereby expressed as absolute values in degrees as well as relative values in percent of the mean value of the first and second measurement sessions. Due to frontal and transverse plane angles close or equal to zero, relative MDCs of the average values are therefore only presented for the sagittal plane. Values in brackets indicate 95 % confidence intervals.

| Activity | Region | Plane | Parameter | Systematic errors[a] | | Relative reliability | Absolute reliability | |
|---|---|---|---|---|---|---|---|---|
| | | | | Mean differences [°] | t-test (p) | ICC | MDC [°] | MDC [%] |
| Standing | Lumbar | Sagittal | Average | −1.9 (-4.1, 0.4) | 0.098 | 0.97 (0.91, 0.99) | 7.3 (3.4, 11.2) | 16.9 (7.9, 25.9) |
| | Thoracic | | | 1.2 (-0.6, 3.0) | 0.166 | 0.89 (0.74, 0.96) | 7.2 (4.1, 10.3) | 20.7 (11.9, 29.5) |
| Sitting | Lumbar | Sagittal | Average | −1.3 (-6.0, 3.5) | 0.554 | 0.80 (0.35, 0.95) | 12.1 (3.9, 20.4) | 59.8 (19.2, 100.4) |
| | Thoracic | | | 1.2 (-0.9, 3.3) | 0.250 | 0.87 (0.69, 0.95) | 8.4 (4.7, 12.1) | 27.4 (15.3, 39.4) |
| Walking | Lumbar | Sagittal | Average | −0.5 (-3.8, 2.8) | 0.740 | 0.88 (0.69, 0.95) | 12.6 (6.9, 18.4) | 29.7 (16.2, 43.2) |
| | | Frontal | | 2.6 (-1.5, 6.7) | 0.191 | 0.31 (-0.19, 0.68) | 15.6 (8.5, 22.6) | – |
| | | Transverse | | 0.7 (-0.5, 1.8) | 0.248 | 0.17 (-0.30, 0.57) | 4.7 (2.7, 6.6) | – |
| | Thoracic | Sagittal | | 0.2 (-1.3, 1.6) | 0.783 | 0.90 (0.76, 0.96) | 5.9 (3.4, 8.4) | 16.0 (9.2, 22.8) |
| | | Frontal | | 0.6 (-1.6, 2.8) | 0.572 | 0.30 (-0.16, 0.66) | 8.8 (5.1, 12.6) | – |
| | | Transverse | | −0.9 (-2.2, 0.4) | 0.177 | 0.52 (0.10, 0.78) | 5.3 (3.0, 7.5) | – |
| | Lumbar | Sagittal | RoM | −1.1 (-3.8, 1.6) | 0.398 | 0.60 (0.18, 0.83) | 10.4 (5.7, 15.1) | 104.8 (57.2, 152.4) |
| | | Frontal | | 2.3 (-0.7, 5.3) | 0.128 | 0.55 (0.11, 0.81) | 11.4 (6.2, 16.6) | 64.1 (35.0, 93.2) |
| | | Transverse | | 0.4 (-0.2, 1.1) | 0.202 | 0.79 (0.53, 0.91) | 2.7 (1.6, 3.9) | 34.2 (19.7, 48.8) |
| | Thoracic | Sagittal | | −1.1 (-2.9, 0.6) | 0.200 | 0.20 (-0.26, 0.59) | 7.1 (4.1, 10.1) | 222.6 (127.8, 317.4) |
| | | Frontal | | 0.5 (-0.7, 1.6) | 0.391 | 0.88 (0.71, 0.95) | 4.6 (2.6, 6.5) | 37.1 (21.3, 52.9) |
| | | Transverse | | −0.5 (-1.7, 0.7) | 0.385 | 0.84 (0.63, 0.94) | 4.9 (2.8, 7.0) | 42.2 (24.2, 60.2) |
| Running | Lumbar | Sagittal | Average | −0.4 (-3.6, 2.8) | 0.792 | 0.89 (0.72, 0.96) | 12.1 (6.6, 17.7) | 30.0 (16.4, 43.6) |
| | | Frontal | | 3.1 (-1.3, 7.5) | 0.150 | 0.16 (-0.33, 0.58) | 16.8 (9.1, 24.4) | – |
| | | Transverse | | 0.5 (-0.9, 1.8) | 0.468 | 0.11 (-0.36, 0.54) | 5.4 (3.0, 7.7) | – |
| | Thoracic | Sagittal | | 0.0 (-1.2, 1.1) | 0.931 | 0.96 (0.89, 0.98) | 4.6 (2.6, 6.6) | 13.7 (7.7, 19.7) |
| | | Frontal | | 1.1 (-1.2, 3.4) | 0.316 | 0.35 (-0.13, 0.69) | 9.0 (5.1, 13.0) | – |
| | | Transverse | | −0.8 (-2.0, 0.3) | 0.148 | 0.55 (0.13, 0.80) | 4.6 (2.6, 6.6) | – |
| | Lumbar | Sagittal | RoM | 1.1 (-1.8, 3.9) | 0.439 | 0.71 (0.36, 0.88) | 10.9 (6.0, 15.9) | 83.3 (45.4, 121.1) |
| | | Frontal | | 0.2 (-1.2, 1.7) | 0.746 | 0.76 (0.45, 0.90) | 5.5 (3.0, 8.0) | 43.9 (24.0, 63.9) |
| | | Transverse | | −0.3 (-1.5, 1.0) | 0.653 | 0.53 (0.09, 0.79) | 4.9 (2.8, 7.1) | 43.8 (24.6, 63.1) |
| | Thoracic | Sagittal | | −0.1 (-0.6, 0.5) | 0.815 | 0.86 (0.66, 0.94) | 2.3 (1.3, 3.2) | 45.0 (25.2, 64.8) |
| | | Frontal | | −0.1 (-1.5, 1.4) | 0.922 | 0.95 (0.88, 0.98) | 5.6 (3.1, 8.0) | 25.2 (14.1, 36.3) |
| | | Transverse | | −0.7 (-2.2, 0.8) | 0.317 | 0.63 (0.24, 0.84) | 5.9 (3.3, 8.4) | 41.0 (23.0, 59.0) |
| Chair up | Lumbar | Sagittal | RoM | −0.8 (-3.6, 2.0) | 0.551 | 0.75 (0.44, 0.90) | 11.2 (6.3, 16.1) | 39.8 (22.3, 57.3) |
| | Thoracic | | | 0.5 (-1.0, 2.0) | 0.486 | 0.43 (-0.01, 0.74) | 6.2 (3.6, 8.9) | 89.0 (51.1, 126.9) |
| Chair down | Lumbar | Sagittal | RoM | 2.5 (0.3, 4.6) | 0.025* | 0.88 (0.71, 0.95) | 8.3 (4.7, 12.0) | 30.3 (17.0, 43.6) |
| | Thoracic | | | 0.5 (-0.9, 1.8) | 0.464 | 0.59 (0.19, 0.82) | 5.5 (3.2, 7.9) | 81.5 (46.8, 116.2) |
| Box up | Lumbar | Sagittal | RoM | 1.1 (-2.6, 4.7) | 0.551 | 0.72 (0.41, 0.88) | 14.8 (8.5, 21.1) | 41.8 (24.0, 59.6) |
| | Thoracic | | | 0.1 (-1.5, 1.8) | 0.860 | 0.33 (-0.13, 0.68) | 6.6 (3.8, 9.4) | 124.3 (71.4, 177.3) |
| Box down | Lumbar | Sagittal | RoM | 0.1 (-4.6, 4.9) | 0.949 | 0.54 (0.13, 0.79) | 19.2 (11.0, 27.4) | 54.7 (31.4, 78.0) |
| | Thoracic | | | −0.2 (-1.9, 1.4) | 0.765 | 0.37 (-0.09, 0.70) | 6.6 (3.8, 9.4) | 115.0 (66.0, 164.0) |
| CMJ | Lumbar | Sagittal | RoM | 0.1 (-4.9, 5.2) | 0.966 | 0.63 (0.21, 0.85) | 18.6 (9.8, 27.3) | 47.6 (25.2, 70.0) |
| | Thoracic | | | −0.1 (-2.6, 2.5) | 0.959 | 0.50 (0.07, 0.77) | 10.5 (6.0, 15) | 81.5 (46.8, 116.2) |

The asterisks (*) indicates a statistically significant systematic error at a p ≤ 0.05 level.

### 3.1. Continuous data

No systematic errors were found (Figs. 1 and 2). ICC values for sagittal plane angles indicated high overall consistency (ICCs mostly >0.75), whereas values for frontal and transverse plane angles during walking and running showed only low to moderate consistency (ICCs mostly <0.6). MDCs for sagittal plane angles ranged from 3.9° to 22.9°, with relative values of <40 % for all thoracic angles as well as lumbar angles during walking and running. Relative MDCs for chair rising and CMJ were <200 % and reached peak values of >300 % during box lifting. Frontal and transverse plane angles showed MDCs of 4.3°-20.8°, which were not expressed as relative values due to multiple angle zero-crossings during movement.

### 3.2. Discrete parameters

No systematic errors were found, except for lumbar RoM during sitting down on a chair (mean difference = 2.5°; p = 0.025) (Table 1). Most ICCs ranged from 0.52 to 0.96, except from a few frontal and transverse plane average angles during walking and running (ICCs of 0.11−0.35) as well as thoracic sagittal plane RoM values during walking, chair rising and box lifting (ICCs of 0.2−0.43). MDCs for sagittal plane angle average and RoM values were 4.6°-13.3° and 2.3°-19.2°, with relative values of <60 % and <230 %, respectively. Frontal and transverse plane angles showed MDCs of 4.6°-16.8° for average (not expressed as relative MDCs due to angles close or equal to zero) and 2.7°-11.4° (relative MDCs <70 %) for RoM values.

### 4. Discussion

This study aimed at quantifying the between-session reliability of continuous and discrete three-dimensional spinal curvature and rotation angles during various functional activities. The analysis indicated high relative consistency for sagittal plane angles during all activities, but not for frontal and transverse plane angles during walking and running. In terms of absolute reliability, MDCs among several angles and activities were mostly below 15°, with relative values ranging between 10 % and 750 %.

Compared to the literature, our results slightly differ from previously published findings [9–12]. However, it should be noted that our marker configuration and definition of kinematic parameters were substantially different, which complicates appropriate direct comparisons.

Despite the possibility that frontal and transverse plane angles during walking and running might simply be less reliable than sagittal plane angles, the lower relative consistency might be partially explained by the considerably smaller movement extent. It was shown that low levels of between-subjects variability can depress ICCs even if the differences between the sessions are low [16]. Momentary ICC fluctuations such as observed during the chair rising, box lifting and CMJ activities, on the other hand, might be more likely ascribed to movement standardization and event detection issues.

Regarding absolute reliability, we did not intend to rate the reported MDCs as high or low, but rather to provide a solid database for choosing the appropriate activities/angles/parameters when planning on investigating longitudinal effects of certain treatment modalities.

An important limitation of this study is that our motion capture system was not always able to identify and distinguish all lumbar spine markers, especially in smaller individuals and individuals with a pronounced lumbar lordosis. We therefore suggest that future studies include more cameras and/or cameras with a higher resolution to overcome such issues. Moreover, the current findings are based on a relatively small sample of asymptomatic healthy individuals, which might limit its generalizability as well as the transferability to patient populations.

This study provides important information that can serve as a basis for researchers and clinicians aiming at investigating longitudinal effects of treatment interventions on spinal motion behavior in patients with spinal pathologies.

**Declaration of Competing Interest**

The authors report no declarations of interest.

**Acknowledgements**

The authors thank Jana Frangi, Edwige Simonet and Magdalena Suter for data collection. This study was partially founded by the Swiss Physiotherapy Association (physioswiss).

**Appendix A. Supplementary data**

Supplementary material related to this article can be found, in the online version, at doi:https://doi.org/10.1016/j.gaitpost.2021.02.008.

**References**

[1] S. Schmid, B. Bruhin, D. Ignasiak, J. Romkes, W.R. Taylor, S.J. Ferguson, et al., Spinal kinematics during gait in healthy individuals across different age groups, Hum. Mov. Sci. 54 (2017) 73–81.
[2] L. Chiari, U. Della Croce, A. Leardini, A. Cappozzo, Human movement analysis using stereophotogrammetry. Part 2: instrumental errors, Gait Posture 21 (2005) 197–211.
[3] U. Della Croce, A. Leardini, L. Chiari, A. Cappozzo, Human movement analysis using stereophotogrammetry. Part 4: assessment of anatomical landmark misplacement and its effects on joint kinematics, Gait Posture 21 (2005) 226–237.
[4] A. Leardini, L. Chiari, U. Della Croce, A. Cappozzo, Human movement analysis using stereophotogrammetry. Part 3. Soft tissue artifact assessment and compensation, Gait Posture 21 (2005) 212–225.
[5] P. Merriaux, Y. Dupuis, R. Boutteau, P. Vasseur, X. Savatier, A study of vicon system positioning performance, Sensors (Basel). (2017) 17.
[6] P. Eichelberger, M. Ferraro, U. Minder, T. Denton, A. Blasimann, F. Krause, et al., Analysis of accuracy in optical motion capture - A protocol for laboratory setup evaluation, J. Biomech. 49 (2016) 2085–2088.
[7] U. della Croce, A. Cappozzo, D.C. Kerrigan, Pelvis and lower limb anatomical landmark calibration precision and its propagation to bone geometry and joint angles, Med. Biol. Eng. Comput. 37 (1999) 155–161.
[8] R. Zemp, R. List, T. Gulay, J.P. Elsig, J. Naxera, W.R. Taylor, et al., Soft tissue artefacts of the human back: comparison of the sagittal curvature of the spine measured using skin markers and an open upright MRI, PLoS One 9 (2014), e95426.
[9] R. Fernandes, P. Armada-da-Silva, A.L. Pool-Goudzwaard, V. Moniz-Pereira, A. P. Veloso, Three dimensional multi-segmental trunk kinematics and kinetics during gait: test-retest reliability and minimal detectable change, Gait Posture 46 (2016) 18–25.
[10] J.M. Wilken, K.M. Rodriguez, M. Brawner, B.J. Darter, Reliability and Minimal Detectible Change values for gait kinematics and kinetics in healthy adults, Gait Posture 35 (2012) 301–307.
[11] F.M. Rast, E.S. Graf, A. Meichtry, J. Kool, C.M. Bauer, Between-day reliability of three-dimensional motion analysis of the trunk: a comparison of marker based protocols, J. Biomech. 49 (2016) 807–811.
[12] D. Sanchez-Zuriaga, J. Lopez-Pascual, D. Garrido-Jaen, M.F. de Moya, J. Prat-Pastor, Reliability and validity of a new objective tool for low back pain functional assessment, Spine. 36 (2011) 1279–1288.
[13] M. Suter, P. Eichelberger, J. Frangi, E. Simonet, H. Baur, S. Schmid, Measuring lumbar back motion during functional activities using a portable strain gauge sensor-based system: A comparative evaluation and reliability study, J. Biomech. 100 (2020), 109593.
[14] S. Schmid, D. Studer, C.C. Hasler, J. Romkes, W.R. Taylor, R. Brunner, et al., Using skin markers for spinal curvature quantification in main thoracic adolescent idiopathic scoliosis: an explorative radiographic study, PLoS One 10 (2015), e0135689.
[15] C. Bangerter, J. Romkes, S. Lorenzetti, A.H. Krieg, C.C. Hasler, R. Brunner, et al., What are the biomechanical consequences of a structural leg length discrepancy on the adolescent spine during walking? Gait Posture 68 (2019) 506–513.
[16] J.P. Weir, Quantifying test-retest reliability using the intraclass correlation coefficient and the SEM, J. Strength Cond. Res. 19 (2005) 231–240.
[17] T.C. Pataky, One-dimensional statistical parametric mapping in Python, Comput. Methods Biomech. Biomed. Engin. 15 (2012) 295–301.
[18] J.M. Bland, D.G. Altman, Applying the right statistics: analyses of measurement studies, Ultrasound Obstet. Gynecol. 22 (2003) 85–93.